\documentclass[aps,prl,showpacs,twocolumn,superscriptaddress,10pt]{revtex4-1}
\usepackage{color}
\usepackage{soul}
\usepackage{amsfonts,amssymb,bm}
\usepackage{amsmath}
\usepackage{graphicx}
\usepackage{epsfig}
\usepackage{epstopdf}

\begin{document}

\title{Identification of Decoherence-Free Subspaces Without Quantum Process Tomography}
\author{D.H. Mahler}
\email[]{dmahler@physics.utoronto.ca}
\author{L. Rozema}
\author{A. Darabi}
\author{A.M. Steinberg}

\affiliation{Centre for Quantum Information \& Quantum Control and Institute for Optical Sciences, Dept. of Physics, 60 St. George St.,
University of Toronto, Toronto, Ontario, Canada M5S 1A7}

\date{\today}

\begin{abstract}
Characterizing a quantum process is the critical first step towards applying such a process in a quantum information protocol.  Full process characterization is known to be extremely resource-intensive, motivating the search for more efficient ways to extract salient information about the process.  An example is the identification of ``decoherence-free subspaces'', in which computation or communications may be carried out, immune to the principal sources of decoherence in the system.  Here we propose and demonstrate a protocol which enables one to directly identify a DFS without carrying out a full reconstruction.  Our protocol offers an up-to-quadratic speedup over standard process tomography.  In this paper, we experimentally identify the DFS of a two-qubit process with 32 measurements rather than the usual 256, characterize the robustness and efficiency of the protocol, and discuss its extension to higher-dimensional systems.

\end{abstract}

\pacs{42.50.Xa,03.65.Wj,03.67.-a}

\maketitle


\section{}
It is now widely appreciated that the cost of fully characterizing a quantum process (``process tomography'' \cite{NC,DFVJ,aePT}) grows exponentially with the size of the system, often making it intractable in practice.  This has motivated much recent work on partial or approximate characterization of quantum states and processes \cite{paz,flammia,shabani,emerson,whitePT}.  Even such techniques leave an important gap between the job of characterization and the goal of actually determining parameters which could be used to optimize the performance of quantum information tasks using the system.
For instance, it is crucial in quantum information processing to mitigate the effects of decoherence; one way of doing so is to use ``decoherence-free subspaces'' \cite{DFS1,DFS2,reviewDFS,zanardi,zanardi2}, corners of Hilbert space that are inherently free of decoherence.  DFSs have been experimentally demonstrated \cite{DFSe1,DFSe2} and used to improve the performance of quantum information protocols \cite{DFSp1,DFSp2}.  In principle, DFSs can be determined from the ``superoperator'' which is extracted in process tomography.
Here we instead present a protocol for determining 
the identity of a DFS {\it directly} from experimental data.  This result could be applied immediately to the use of a system for quantum communications or information processing.  Our protocol offers a polynomial (up to quadratic) speedup over full process tomography, and we demonstrate it experimentally, characterizing its performance for a two-photon process.  We succeed in identifying a 3-dimensional DFS in 32 measurements, to be compared with the 256 required for full process tomography.

DFS's are subspaces of the original Hilbert space that are intrinsically free from decoherence.  We define a DFS as follows:  Given a process $\varepsilon (\cdot)$, a subspace is decoherence-free if the effect of the process on any state beginning in the DFS is merely a unitary transformation.  Using process tomography, $\varepsilon (\cdot)$ can be characterized completely.  One can view the entire process as a black box, and then feed input states into the box one at a time, after which full state tomography is done on each output state.  Once a sufficient number of states has been sent through, the process matrix can be determined via linear inversion or maximum likelihood techniques.  For an n-qubit operation, the number of measurements needed for full process tomography scales as $16^{n}$.  Some work\cite{paz,flammia,shabani,emerson,whitePT}, both theoretical and experimental, has shown examples of cases in which it is possible to perform these tasks more efficiently,  however, even the density or process matrices produced by such protocols are not directly useable objects, and must be further analyzed to produce useful information.  There has also been progress\cite{walborn, adamson} towards directly measuring quantities such as the purity or the tangle of quantum states, but again, these numbers serve only to characterize imperfections, not to provide a path forwards towards mitigating their deleterious effects.  To this end, we present a method which is efficient and provides a method for avoiding decoherence.

To understand our method for locating a DFS, consider a model 2-qubit system possessing a 3D DFS relevant to some optical experiments: the ``sometimes swap'' (SSWAP), in which every time a 2-qubit state traverses the channel there is a probability (for instance, 50\%) that the states of the two qubits will be swapped.  The process has the form 
\begin{equation}
\varepsilon (\rho) = (1-p)I\rho I + pS\rho S \; ,
\end{equation}
where p is the swap probability and S is the qubit swap operator: $S\vert\psi_1\rangle\vert\psi_2\rangle = \vert\psi_2\rangle\vert\psi_1\rangle$.

  If this channel were classical, the $\vert 00 \rangle$ and $\vert 11 \rangle$ states would be completely insensitive to the error, while the $\vert 01 \rangle$ and $\vert 10 \rangle$ states would be prone to it.  Using quantum mechanics, we can construct a third state that is insensitive to this error, namely $\frac{1}{\sqrt{2}}(\vert 01 \rangle + \vert 10 \rangle)$.  Thus, this channel possesses a 3-dimensional DFS.  Together, these states are the familiar triplet states.  However, the remaining state - the singlet state $\frac{1}{\sqrt{2}}(\vert 01 \rangle - \vert 10 \rangle)$ - receives a $\pi$-phase shift a fraction p of the time and none the fraction 1-p of the time and is thus decohered from the other three states in the set.  Consider the action of this channel, with a swap probability p=0.5, on an arbitrary input state, which will in general have some finite overlap with both the singlet state and the subspace formed by the triplet states:
\begin{equation}
\vert \psi_{in} \rangle = \alpha \vert \psi_{t} \rangle + \beta \vert \psi_{s} \rangle
\end{equation}
where $\vert \psi_{t} \rangle$ is some element of the triplet subspace.  Then,
\begin{eqnarray} 
\rho_{out} &=& \varepsilon (\vert \psi_{in} \rangle\langle \psi_{in} \vert) = \vert \alpha \vert^2 \vert \psi_{t} \rangle \langle \psi_{t} \vert + \vert \beta \vert^2 \vert \psi_{s} \rangle \langle \psi_{s} \vert 
\end{eqnarray}
The two subspaces are completely decohered from each other, meaning that all the coherences between subspaces are zero.  If one then performs state tomography and an eigenvector decomposition on the reconstructed density matrix, one of the eigenvectors is the 1D DFS (the singlet), and the other lies somewhere in the 3D DFS.  The challenge comes in that we do not know which eigenvector belongs to which subspace. 

Sending a second random state through the channel, and comparing the two lists of eigenvectors yields the answer, because only the singlet vector will in general be common to both eigenvector decompositions.  After just $2d^2=32$ measurements the decoherence-free subspaces have been located. 

In a more general case, there may be unitary rotations before and after the decohering process, so that the decoherence-free subspaces are rotated between the input and the output.  Consider, in other words, a channel of the following form:
\begin{equation}
U_2 \varepsilon (U_1 \rho U_1^{\dagger} ) U_2^{\dagger} = \frac{1}{2}( U_2 U_1 \rho U_1^{\dagger}U_2^{\dagger} + U_2 S U_1 \rho U_1^{\dagger} S U_2^{\dagger}) \; .
\end{equation}
The state $U_1^{\dagger} \vert \psi_{s} \rangle$ corresponds to the 1D DFS at the input, and is transformed to $U_2 \vert \psi_{s} \rangle$ at the output, while the 3D subspaces orthogonal to these remain decoherence-free and experience the channel as a unitary rotation.  Our goal is to find the appropriate subspace for encoding at the input.  However,
the eigenvector common to two output density matrices will be $U_2 \vert \psi_{s} \rangle$.  Such a state would not, in general, remain pure after the action of the process. Rather, it is the state which would remain pure if it could be sent \emph{backwards} through the process. To reconstruct the DFSs at the \emph{input} of the process, one might imagine a time-reversed tomography.  Instead of sending a state through the input of the process and measuring observables at the output, one might send states through the output of the process and measure observables at the input.  The action of the process would then be:
\begin{equation}
U_1^{\dagger} \varepsilon (U_2^{\dagger} \rho U_2 ) U_1
\end{equation}
and application of the protocol would yield $U_1^{\dagger} \vert \psi_{s} \rangle$, as desired. 
In many common experimental situations, however, there is a physical distinction between the source and the detector which cannot be feasibly inverted.  A new approach must therefore be identified.

We implement an effective time-reversal of the protocol in the following way: Instead of sending one random state into the input and making $2^{(2n)} =16$ linearly independent projections at the output, we send in 16 linearly independent states at the input and project onto a single random state at the output.  The measurement statistics observed are identical to those which would be observed if a single quantum state could be sent through the process in reverse and measured at the input.  Mathematically, this can be seen from the result
$tr(\rho U_2 \varepsilon(U_1 \hat{O} U_1^{\dagger}) U_2^{\dagger}) = tr(O U_1^{\dagger} \varepsilon(U_2^{\dagger} \rho U_2) U_1)$,
where $\hat{O}$ is one of 16 projectors used in standard tomography.  

Thus, by using these measurements to reconstruct a density matrix and doing an eigenvector decomposition as before, we can reconstruct $U_1^{\dagger} \vert \psi_{s} \rangle$ - the one dimensional DFS at the input - without having any knowledge of $U_1$. This is done in a total of 32 measurements, which is significantly fewer than the 256 measurements required for full process tomography.  

\begin{figure}[ht]
\includegraphics[scale=0.3]{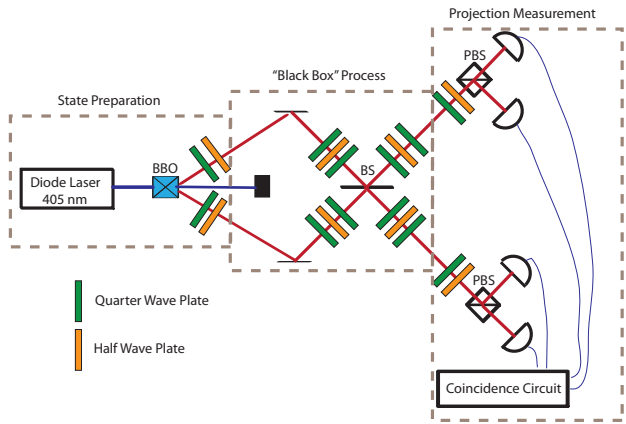}%
\caption{  \label{fig1}  Spontaneous parametric downconversion is performed by pumping a BBO crystal with horizontally polarized light.  A rotation using a quarter-half waveplate combination prepares the two-qubit state.  Next, the unitary rotation $U_1$ is performed using a quarter-half-quarter waveplate combination.  A beamsplitter effects the sometimes-swap gate, after which another quarter-half-quarter waveplate combination is used to perform $U_2$, and finally a quarter-half waveplate combination followed by a polarizing beamsplitter is used to project onto the appropriate state.}
\end{figure}

The sometimes-swap channel was implemented experimentally in a linear-optics set up, shown in figure \ref{fig1}.  Using spontaneous parametric downconversion, photon pairs were created.  Next, using the polarization degree of freedom of the photons, the initial state of the system was set using combinations of quarter and half waveplates in the usual way\cite{JKMW}. 

The decohering gate is realized with a 50/50 beamsplitter and post-selection.  Each photon has a 50\% probability of either being reflected or transmitted, and thus when we postselect on the two photons exiting out of different ports of the beamsplitter it is as if the photons had a 50\% probability of either being swapped or not.  When the path lengths up to the beamsplitter are not equal, the two photons are distinguishable and each photon exits the beamsplitter in a different temporal mode.  When the photons are detected, the temporal mode is traced over and decoherence is produced.  The photons are detected using single photon counting modules.  Tomography on the resulting state of the photons is done using waveplates and polarization beamsplitters.  
To test the protocol, first a separable state was randomly chosen to project onto.  Then the 16 linearly independent states from \cite{JKMW} were sent into the input.  From the measurement outcomes, a density matrix was reconstructed using a maximum likelihood state reconstruction similar to \cite{JKMW}, using efficient convex optimization solvers\cite{yalmip}, and eigenvector decomposition was performed.  This was done eleven times.   Each of the 55 pairs formed by these 11 states is considered one trial of the protocol.  In each trial, the pair of eigenvectors with the highest fidelity was selected as belonging to the 1D DFS.  Since there is measurement noise present in the system (not to be confused with the decoherence which we are trying to characterize), the pairs of eigenvectors are not identical as they are in the idealized theory.  However, if the noise is isotropic, each pair can be ``averaged'' to obtain a better estimate of the 1D DFS.  We define the ``average'' of two pure states as the eigenvector of $\rho_{av}= (\rho_1+\rho_2)/2$ with the largest eigenvector.

\begin{figure}[ht]
\includegraphics[scale=0.5]{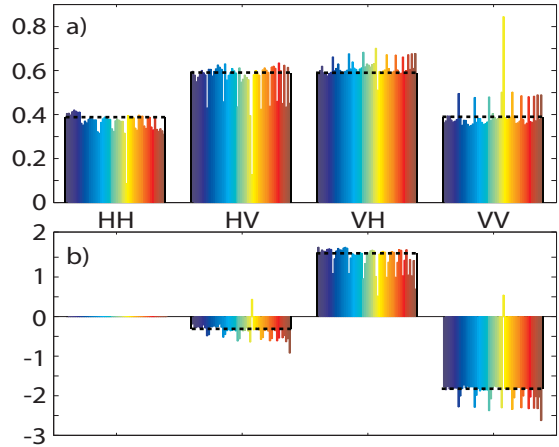}%
\caption{  \label{figdat} The a) amplitudes and b) phases of the reconstructed 1D DFS for 55 random trials of the protocol (each trial plotted as a different colour).  The result of full process tomography is plotted as a black dotted line.}
\end{figure}

   The results of these 55 trials of the protocol are plotted in figure \ref{figdat}.  For 11 of these trials, the estimate for the 1D DFS has a fidelity with the true 1D DFS which is less than expected from noiseless simulations. Ten of these trials' poor performance can be explained as follows:  When the random state chosen has nearly equal projections onto the two DFSs, the density matrix after the process is identity over the subspace formed by its two eigenvectors and its eigenvector decomposition is not unique.  Thus, states that are close to spanning the two DFSs equally are more sensitive to any noise present in the system.  In this case, however, it is possible to simply discard these results, since equal eigenvalues serve as an indicator of the state's unreliability.  In our case, there was one state whose two largest eigenvalues had a ratio of 1.08, while the other ten ratios ranged from 1.64 to 71.5; if the state which had roughly equal eigenvalues is rejected, 10 of the offending trials are rejected as well.  The one other trial whose performance is surprisingly poor can be explained as well.  The overlap between our presumed singlet-state vectors is never exactly unity, and the overlap between two randomly chosen triplet-state vectors may happen to be large.  In the 11th poor trial, the overlap of the two states which were approximately equal to the singlet was 95\%, while the overlap of the two triplet states happened to be 99\%.  Excluding the 10 cases we could reject based on the equal eigenvalues, none of the other 44 trials had a second-largest overlap greater than 3\%.  Thus, all the errors of our protocol can be attributed to uncommon syndromes which are easily detectable, affording the experimenter an opportunity to send in a third state and obtain a reliable estimate in cases where these errors arise.

If the process is completely decohering, all coherences between decoherence-free subspaces decay to zero by the end of the interaction and eigenvector decomposition yields their location.  Complete decoherence is equivalent to the swap probability being 50\%.  When the swap probability is not 50\%, some coherence is left between subspaces, and eigenvector decomposition will not yield the exact location of the DFS.   The swap probability was measured to be 0.51.  Swap probabilities different from 0.5 can be studied by azimuthally misaligning the beamsplitter used to perform the SSWAP.  In this way, the reflected beams can be misaligned while leaving the transmitted beams unaffected. This reduction in the collection efficiency for the reflected photons effectively amounts to an adjustment of the reflectance and transmission probabilities.  

\begin{figure}[ht]
\includegraphics[scale=0.455]{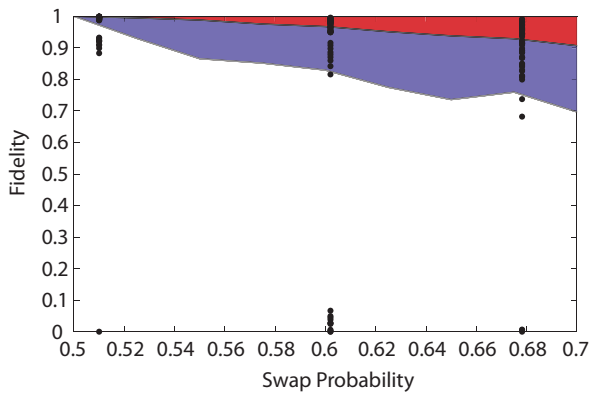}%
\caption{  \label{fig4} The experimental fidelity of the one-dimensional DFS constructed using the protocol with the DFS extracted from full process tomography, as a function of the swap probability for 55 trials of the protocol.  The $63\%$ and $95\%$ confidence regions obtained from noiseless simulation are shaded in red and blue, respectively. }
\end{figure}

Using this technique, the experiment was performed several times for different swap probabilities.  The results are plotted in figure \ref{fig4}, with the confidence regions containing 63\% and 95\% of the results from noiseless simulations plotted in red and blue, respectively. (The experimental points which fall outside these ranges arise due to noise, and can be explained -- or rejected -- based on the criteria described earlier.) One can see that as the decoherence is made less complete (a swap probability further from 0.5), our ability to find the decoherence-free subspaces is hindered.  Simulations show that for any swap probability, the distribution of fidelities of the 1D DFS determined by the protocol with the true 1D DFS has two peaks.  The first occurs at a fidelity of 1, meaning the most likely result is that the protocol results in a high fidelity identification of the DFSs, and the second at a fidelity of 0 (with a much smaller amplitude).  The second part of the distribution, which is peaked at a fidelity of 0, is the result of the protocol choosing the wrong pair of eigenvectors, as described before.  For strong decoherence (swap probability $~50\%$), the probability of this occuring goes as approximately $O\left(\frac{1}{\sqrt{N}}\right)$, where N is the number iterations of each measurement.  See \cite{SM} for details.  

In a second experiment we tested our results by preparing various states in the DFSs identified by our protocol, sening them through the channel, and directly measuring the purity of the output states; we did this for a range of different swap probabilities.  For the 51\% reflectivity, using the channel our protocol identified increased the average purity from 65\% to about 98\%; even for reflectivities of 59\% and 72\%, we consistently achieved purities of approximately 90\%; see \cite{SM} for details.

Finally, consider an even more general case where nothing is known about the channel.  A one dimensional DFS may not exist, making easy characterization impossible. To generalize the discussion to more complicated processes, we begin with the previous treatment.  In the ideal scenario, $d^2$ linearly independent states are sent through a process, and a random state is projected onto at the output.  These measurements are used to reconstruct a density matrix, and eigenvector decomposition is performed.  From just these $d^2$ measurements, something can be learned.  If, for example, there are $d$ eigenvectors with a non-zero eigenvalue, already it is known that no DFS of dimension greater than one exists, and there is no qubit that will remain decoherence-free.  If, however, the number of eigenvectors with non-zero eigenvalues is less than the dimension of the system, a DFS may exist and we must make another set of measurements to learn more.  After each set of $d^2$ measurements, a list of eigenvectors is obtained.  Each eigenvector belongs to a subspace of unknown dimension that is orthogonal to all the other eigenvector subspaces.  In this way, the eigenvectors can be arranged into groups, and Gram-Schmidt orthogonalization can be used to reconstruct the set of subspaces.  Complete knowledge of a subspace is gained once Gram-Schmidt orthogonalization yields the null vector.  The protocol terminates once this has happened for all but one of the subspaces, the last of which can be inferred.  In the presence of finite errors, this protocol could be implimented to yield high-fidelity estimates of DFSs by introducing a threshold for the length of the vector obtained by Gram-Schmidt orthogonalization, below which the protocol would terminate, and an averaging procedure similar to the previously described method.    

In addition to being robust to deviations from complete decoherence, this protocol scales well.  In a system of dimension $d$, each state tomography requires $d^2$ measurements.  In the worst-case scenario, the state space is broken into two decoherence-free subspaces of approximately equal dimension, $\frac{d}{2}$.  In this case, a total of $\frac{d}{2}$ basis vectors will be needed to reconstruct the spaces using the Gram-Schmidt method.   This amounts to $\frac{d}{2}$ states, each of which requires $d^2$ measurements to characterize, for a total of O($d^3$) measurements.  This is a factor of $d$ better than standard process tomography.  It should be noted that this is a worst case scenario, and the protocol can perform even better for different DFS dimensionalities.  For example, when the state space possesses a 1D DFS and a d-1 dimensional DFS, only $2d^2$ measurements are needed.  Thus, the protocol provides a quadratic speedup in the best case.

In conclusion, we have shown that it is possible to experimentally measure characteristics of a process which can be used directly for determining how best to incorporate the process into a larger quantum information system, far more efficiently than through full process tomography.  Specifically, our protocol enables efficient, direct identification and characterization of decoherence-free subspaces.  We have used the protocol to measure the identity of a DFS, and then characterized the average purity of this subspace.  The algorithm provides a polynomial speedup compared to standard process tomography schemes.  The protocol requires no ancillary qubits or highly entangled states to be prepared, and only requires simple single-qubit tomographic measurements to be made on those states.  This algorithm provides a method for directly measuring DFSs, without the resources required for full process tomography.  We believe that it and partial-characterization protocols like it will prove essential to the development of quantum technologies in higher-dimensional Hilbert spaces.

\begin{acknowledgments}
We thank NSERC, CIFAR, and QuantumWorks for support.  Additional thanks go to Rob Adamson and Morgan Mitchell for input and ideas, as well as Alan Stummer for designing the coincidence circuit and Robin Blume-Kohout for interesting conversations.
\end{acknowledgments}


\section{Supplementary Material}

\subsection{Direct Measurement of DFS Purity}

\begin{figure}[ht!]
\includegraphics[scale=0.3]{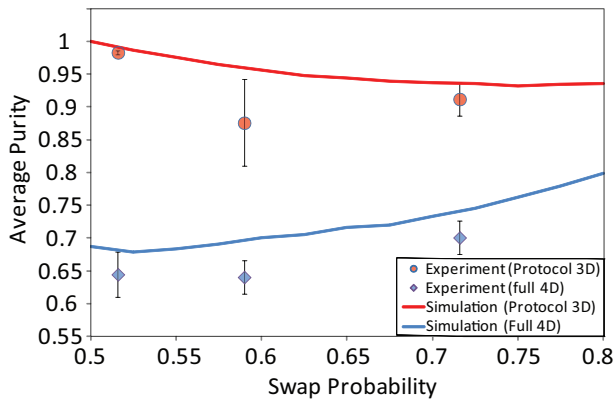}%
\caption{  \label{figdat2} The measured average purities over separable states of the 3D DFSs predicted by the protocol and the measured average purity over all separable states in the 4D Hilbert space, as a function of the swap probability.  The result of simulations is plotted as solid lines. }
\end{figure}

An experimentally relevant figure of merit is the average purity of the DFSs predicted by our protocol.  The average purity of the subspace will not only tell us how close the subspace is to a true DFS, but also how useful the subspace might be for transmitting pure states, in comparison with the entire 4D Hilbert space that we have access to.  For different values of the swap probability, several trials of our protocol are performed, without averaging:  three states are chosen to project onto, and from their eigenvectors the one closest to the 1D DFS is chosen.  A subspace formed by the remaining three eigenvectors is constructed, and then using these results the average purity of the 3D DFS is experimentally measured by sending in many states belonging to that subspace through the process and performing state tomography at the output.  The purities of these states are then averaged, and the whole process repeated for several different swap probabilities to produce figure \ref{figdat2}.  As the swap probability is increased from $50\%$ the DFSs predicted by the protocol are no longer exact, but the decoherence becomes smaller and the average purity remains high.  Even for smaller amounts of decoherence one can see that a significant gain in average purity can be achieved by making use of the reduced-decoherence subspaces identified by our protocol.  It should also be noted that for large deviations from complete decoherence, the decoherence has almost no effect on the system anyway.  When the swap probability is very low, the swap almost never occurs and decoherence is not observed.  When the swap probability is very large, the swap almost always occurs, and the process is very close to being unitary.

\subsection{Error Syndrome Frequency and Detection}
As mentioned in the body of the paper, in the presence of finite noise, the protocol may misidentify DFSs: for the case where we are comparing a random state from a 3D Hilbert space to two singlet vectors that should be identical but are not due to noise, failure occurs when the fidelity between the two random triplet vectors is higher than the fidelity between the two noisy singlets.  If N copies of the quantum state are used to perform state tomography, it is reasonable to assume that the fidelity between two quantum states is approximately $1-\frac{1}{\sqrt{N}}$.  Also, it is well known \cite{funny} that in D dimensions the fidelity of two random pure states with each other follows an exponential distribution in the limit of large D \cite{WootersExp}:
\begin{equation}
P(F)=\frac{e^{-FD}}{\int_0^1{e^{-FD}dF}}
\end{equation} 

Thus, the failure probability, which corresponds to the probability that the two random triplet states ($\vert \psi_{tr1}$ and $\vert \psi_{tr2}$) have a higher fidelity than the singlet vectors (which will have an overlap $O\left(1-\frac{1}{\sqrt{N}}\right)$) is:

\begin{eqnarray}
P(F(\vert \psi_{tr1},\vert \psi_{tr2}) &>& 1-\frac{1}{\sqrt{N}}) \nonumber \\
&=& \int_{1-\frac{1}{\sqrt{N}}}^{1}P(F) \nonumber \\
&=& \frac{e^{\frac{d}{\sqrt{N}}}-1}{e^{d}-1} \approx O(\frac{1}{\sqrt{N}})
\end{eqnarray}

\begin{figure}[ht!]
\includegraphics[scale=0.36]{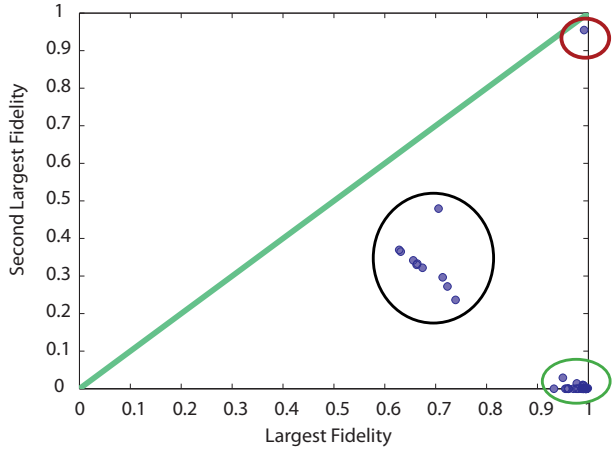}%
\caption{  \label{figL} The fidelity of the largest fidelity pair of eigenvectors vs. the fidelity of the second-largest fidelity pair of eigenvectors, for all 55 trials of the protocol.  The trials with poor performance are circled in red and black, while the high performance trials are circled in green. }
\end{figure}

This failure mode can be represented graphically and subsequently identified by, when comparing the two lists of eigenvectors to determine which pair has the largest fidelity, plotting the largest fidelity vs. the second largest fidelity.  If the largest fidelity and second largest fidelity are similar, it is an indication that the trial should not be trusted.  In figure \ref{figL}, the fidelity for the largest fidelity pair of eigenvectors is plotted vs. the fidelity of the second-largest fidelity, for a swap probability of 0.51.  The closer these points lie to the green line, which is simply the line of equal fidelities, the less trusted the trial is.  One can see inside the red circle lies the trial which failed with a fidelity of nearly zero.  In the black circle are the trials which were below average performance.  Finally, within the green circle lies the remaining 44 trials which were seen to succeed.

\end{document}